**Coupling Dynamics and Linear Polarization Phenomena in Codirectional Polariton Waveguide Couplers**

Elena Rozas[1], Alexey Yulin[2], Sebastian Klembt[3], Sven Höfling[3], María Dolores Martín[1,4,a)], and Luis Viña[1,4]

[1]*Departamento de Física de Materiales and Instituto Nicolás Cabrera, Universidad Autónoma de Madrid, 28049 Madrid, Spain.*
[2]*Department of Physics and Engineering, ITMO University, St. Petersburg 197101, Russia.*
[3]*Technische Physik, Wilhelm-Conrad-Röntgen-Research Center for Complex Material Systems, and Würzburg-Dresden Cluster of Excellence ct.qmat, Universität Würzburg, D-97074 Würzburg, Germany.*
[4]*Instituto de Física de la Materia Condensada, Universidad Autónoma de Madrid, 28049 Madrid, Spain*
[a)] *Author to whom correspondence should be addressed: dolores.martin@uam.es*

**ABSTRACT**

In this work, we extend our previous studies investigating the potential of integrated devices using polariton waveguides that form codirectional couplers. For suitable coupler parameters, a transfer of condensates between the arms of the coupler, occurs leading to the observation of previously predicted Josephson-like oscillations. The ability to tune the periodicity of these oscillations opens the way to the design of polaritonic circuits in which the directionality of the signal towards the output terminals can be controlled, for a fixed separation between the arms, varying the coupling length. We also investigate the response of the devices to linearly polarized excitation, delving into the dynamics of linear polarization at the output terminals of long couplers, providing valuable insights into their potential applications, including polariton switches and logic gates with efficient operation. Our results are supported by numerical simulations based on the generalized Gross-Pitaevskii equation describing the dynamics of coherent polaritons in spatially non-uniform systems, reproducing the polariton distribution on the output terminals and an oscillatory dependence of the corresponding emission as a function of the polarization azimuth. We show, how the coupling and the controllable polarization degree of freedom in polariton couplers opens avenues for innovative optical architectures and functionalities.

**I. INTRODUCTION**

In recent decades, there has been a growing interest in enhancing the speed and performance of conventional processors, leading to the development of novel photon-based devices [1–3]. This trend is driven by the need for faster operation speeds of processors and the advantages of transferring and processing all-optical information [1, 4]. To meet these demands, researchers have focused on efficiently confining and manipulating light within specific spatial regions. One of the approaches to do so is exploiting semiconductor microcavities, where a Bose – Einstein like condensation (BEC) of exciton – polaritons can be achieved [5-8]. Furthermore, the one-dimensional propagation over very long distances of these BEC condensates has been demonstrated [9], resulting in the realization of various all-optical architectures based on microcavity polaritons [10–18].

The construction of integral components for guiding polaritons to different spatial regions in a sample involves the creation of 1D planar straight waveguides. These waveguides have been extensively discussed in the literature and are commonly employed in polaritonic devices [19–22]. When subjected to high non-resonant excitation power densities, polaritons within waveguides acquire a significant wavevector, predominantly in one component, enabling them to propagate over long distances throughout their lifetime. However, practical devices necessitate more sophisticated structures than simple straight waveguides. By introducing precise defects or modifying the shape of straight waveguides, it becomes possible to guide polaritons along new curved trajectories, building for instance a Mach-Zehnder interferometer [23] or couplers where polaritons can be created at the input terminals, manipulated in the coupling area and retrieved at the output terminals as shown in our previous works [17, 24, 25]. The overall losses in bent waveguides play a crucial role in determining the optical efficiency and the feasibility of large-scale optical devices. A reduction of these losses could be obtained by using e.g. etch-and-overgrowth designs as in Refs. 26 and 27. A noteworthy distinction between conventional waveguides used for light propagation and polariton waveguides is the capability of transitioning from single-mode to multi-mode operation, allowing for the presence of different propagating modes [27-29]. Polariton propagation is studied not only in conventional semiconductor waveguides (GaAs, GaN) but also in systems such as 2D materials [30-32] and perovskites [33-36], where the excitons have very large oscillator strengths. It is also worth mentioning those systems, such as slab waveguides, for which the strong exciton-photon coupling occurs for relatively large wavevectors, forming polaritons with large group velocity and, correspondingly, long propagation length. Linear and nonlinear propagation of polaritons, including solitons, in such systems have been extensively studied in the literature [32,37-39].

The controllability of polariton spin through the polarization state of the excitation light adds another compelling aspect to their properties. This characteristic presents an opportunity to explore a novel category of spin-based integrated devices, which hold great promise for applications including polariton switches, logic gates, and quantum computing, with efficient and ultrafast operation [40–42]. Experimental studies have demonstrated that the interaction between polaritons, dependent on their spins, allows for the creation of such devices, where the output spin state can be manipulated by the polarization of the excitation light [43–45].

As a first approximation, to investigate the potential utilization of the spin degree of freedom in a coupler device, since a circular polarization, directly related to the polariton spin, can be decomposed into a superposition of linear polarizations, we examine its response to linearly polarized non-resonant excitation, which enables the creation of propagating condensates [9]. The characteristics of the polaritons in these couplers and specifically their dispersions are shown below (above) the condensation threshold in figures 2 and 3 (4 and 5) of Ref. 24. Specifically, we focus on couplers featuring narrower waveguides, with a width of 2 µm, while varying the separation between the guides from 0.2 µm to 0.5 µm. This range of parameters facilitates the transfer of condensates between the arms of the coupler, enabling the observation of Josephson-like oscillations [25]; as a step forward to this work, we exploit the periodicity of these oscillations in combination with a variation of the coupling length to select the polariton distribution along a specific output terminal. Additionally, we delve into the analysis of linear polarization dynamics at the output terminals of unexplored long couplers, which facilitate the interaction between the eigenstates of linearly polarized polaritons along the coupling length. This interaction results in an oscillatory behavior as a function of the analyzer angle of the polariton distribution at the output terminals, which determines the relative phase between the signal measured at the left- and right-output terminals, as borne out from our simulations.

The Gross-Pitaevskii equation (GPE) has proven to be a powerful tool for describing the dynamics of the polariton condensates in the mean field approximation [46, 47]. In order to reproduce both the creation of the coherent polaritons within the excitation spot and their propagation along a waveguide, the equation with an incoherent pump (non-resonant excitation) must be used. Incoherent pump means that there is a linear gain in the system due to the incoherent exciton reservoir created by the pump. However, obtaining a quantitative agreement between model and experiments is a challenging task, as the model includes many coefficients that are only approximately known. In addition, these parameters strongly affect the polaritonic dynamics. This problem is usually solved by a careful fine-tuning of the parameters. In contrast, the analysis of polariton propagation is a much simpler task because it usually occurs in a quasilinear regime, which reduces the number of physical processes affecting the polariton dynamics and thus eases the requirements for the accuracy of the parameters.

Since the main goal of this work is to study the propagation of polaritons in waveguide systems, we chose to replace the incoherent pump by a coherent one (resonant excitation). The pump characteristics can be easily adjusted to create polaritons with properties corresponding to those observed experimentally. In the Supporting Information we show that the incoherent pump approach gives results that are in qualitative agreement with those obtained by the simpler model with the coherent pump. We also discuss there the differences between the results obtained by the simulations performed with both models. With our modeling we explain novel experimental results in microcavity couplers.

## II. SAMPLES AND EXPERIMENTAL SETUP

The sample consists of numerous directional couplers, namely those used in our previous works [24, 25], with different size parameters in terms of width (w: transversal dimension of the waveguide), coupling length ($L_c$: extent where waveguides are parallel) and distance between the arms (d) (Figure 1). To build these couplers we start with a MBE grown two-dimensional cavity built with two DBR mirrors, consisting of 23/27 pairs of alternating layers of $Al_{0.2}Ga_{0.8}As$/AlAs in the top/bottom. In the antinode positions of the electromagnetic field confined between these mirrors we have placed three sets of 4 GaAs quantum wells (QWs), 7 nm wide, separated by 4 nm AlAs barriers. A Q-factor exceeding 5000 has been determined experimentally by low power photoluminescence measurements for a large negative detuning of the lower polariton branch. From its Lorentzian linewidth, ~0.3 meV, we directly infer the resonator's Q-factor. Then, the microcavity wafer is processed by reactive ion etching down to the QWs, carving a pattern of adjacent waveguides of varying $L_c$ (from <2 to 100 μm), w (2 or 6 μm), and d (0.2 to 1.5 μm).

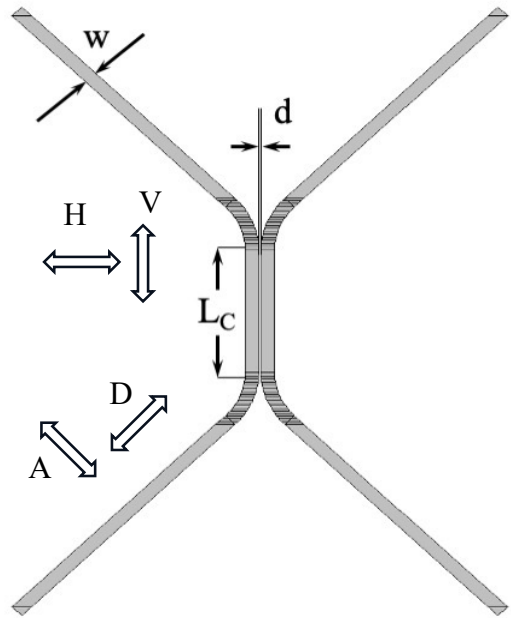

Figure 1.- Sketch of a coupler: input/output terminals on the top/bottom, linked through the coupling length ($L_C$). The separation between the coupler's arms (d) and the coupler's width (w) are also shown.

During the experiments, the couplers are kept in a cold finger cryostat at 12 K. We study the movement of the polariton condensates, along the whole coupler structure by measuring the light (PL) emitted. The condensates are created focusing 2 ps long pulses derived from a Ti:Al$_2$O$_3$ laser, tuned at 745 nm, on the sample's surface with a microscope objective (20x, NA=0.45, spot size 1.5 μm). The same objective is used to collect the PL, whose polarization is analyzed into the different azimuths by a combination of a half-wave plate and a linear polarizer placed in front of the entrance slit of the imaging spectrometer used to resolve it in energy.

## III. RESULTS
### (A) Dynamics of polariton coupling

The couplers are pumped only at one of the input terminals, while the second arm is eventually populated by the tunneling of the condensate wavefunction to the adjacent waveguide. To secure this coupling, the energy states in each waveguide must be resonant and present a similar condensation threshold ($P_{th} \approx 12$ kW/cm$^2$). To ensure that these conditions are met, we use a spatial filter in real space to obtain a clean emission from the different parts of the coupler, blocking the emission originating from selected parts of the coupler. The signal intensity decreases considerably as polaritons propagate along the structure, therefore, emissions with very different intensities are found at each part of the coupler. Blocking the region with the most intense emission allows adjusting the detection conditions to observe weaker signals in other regions, such as the output terminals, where signals 95% less intense than the initially injected signal are found.

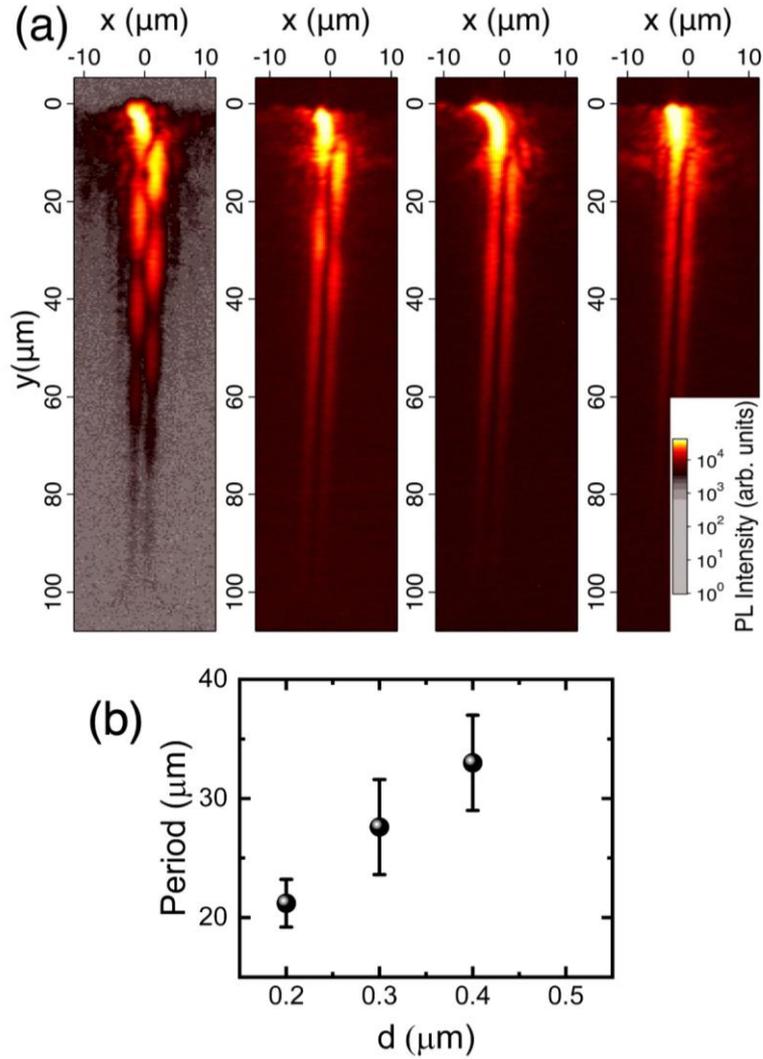

Figure 2.- (a) Real-space PL emission maps along the coupling region of several directional couplers with increasing separation between the arms: from left to right 0.2, 0.3, 0.4 and 0.5 μm. The pumped input terminal is removed from the images using a spatial filter. False color intensity scale is logarithmic. (b) Periodicity of the Josephson-like oscillations along the coupling region as a function of the separation between the coupler's arms.

We first discuss the dependence of the coupling between the two arms as a function of the separation between the waveguides, d. To do so, we block the emission from the input terminal of different couplers with the same size parameters (width and coupling length) but different d. The real-space emission maps of the coupling region of these couplers, at 780.8 nm, are depicted in Figure 2(a), from left to right: d = 0.2, 0.3, 0.4 and 0.5 μm, respectively. All of them are pumped on the filtered-out region at the top-left terminal of the structure, so that y = 0 corresponds to the beginning of the coupling region. A clear exchange of polaritons between the two waveguides is observed. This population transfer is characterized by a strong signal oscillating between the two arms when they are sufficiently close to each other. The oscillations progressively vanish as the separation d increases, disappearing completely when d = 0.5 μm. These oscillations have been theoretically modelled using a slowly varying amplitude approach for the modes of the coupled waveguides [25] and denominated Josephson-like oscillations [48-50]. The transfer of polaritons from one waveguide to the other gives rise to a dynamical renormalization of the energy in each waveguide, resulting in damped harmonic oscillations. An analysis of the periodicity of the oscillations, shown in Figure 2(b), finds a significant increase of the period as a function of d. This is the coupling strength between the waveguides depends not only on the depth but also on the width of the Josephson-like potential [25]: a wider barrier (larger d) will result in a smaller coupling, thus hindering the tunneling of polaritons from one waveguide to the other and consequently increasing the periodicity. All these facts support the identification of these population beats as Josephson-like oscillations.

The ability to tune the periodicity of these Josephson-like oscillations opens up the way to the design of more sophisticated polaritonic circuits in which the directionality of the signal towards the output terminals can be controlled. Following this idea, we investigate how the signal can be steered to a chosen output terminal by the selection of specific size parameters of the coupler. We consider four devices with w = 2 µm and d = 0.2 µm, in which the coupling length $L_c$ nominally varies from ≲2 to 50 µm, displaying their real space emission maps in Figure 3. All are pumped on the top-left terminal, creating condensates, at 780.8 nm, that travel throughout the coupler until they reach the left (L) or right (R) output terminal. The logarithmic false colour scale is saturated in the pump area to enhance the visibility of weaker signals at the output terminals. In the case of $L_c$ < 2 µm (first panel from the left), the condensate is not able to couple to the other arm because of its short length, thus, the signal is mainly funneled towards the L output terminal. When $L_c$ is increased up to 10 µm (second panel from the left) the signal is coupled to the neighboring arm since the coupling length is shorter than the oscillation period (13 µm for this d). The system is not able to transfer polaritons back to the pumped arm, thus, the signal is mainly driven towards the R output terminal. In the case of $L_c$ = 20 µm (third panel from the left), the signal is guided mostly towards the L terminal: now the coupling length is long enough to allow for a second oscillation of the population, transferring polaritons back to the pumped waveguide. Finally, in the case of $L_c$ = 50 µm a new ratio in the distribution of the signal is found: the emission intensity through both terminals is comparable, obtaining an effective population splitting of ∼ 50%. To summarize these findings, we have obtained the ratio between the signal guided into the L terminal and the total intensity at the output terminals. To do this, we integrate the PL intensity throughout the output terminals finding $I_L$ and $I_R$ ($I_{L/R}$ denotes the integrated signal intensity along the left/right output terminal) and obtain the fraction R= $I_L/(I_L+I_R)$. This quotient, R, is plotted as a function of the coupling length in Figure 4(b): the experimental points (open circles) are obtained from the analysis of the emissions shown in Fig. 3. A drastic change in the direction of the output signal can be clearly observed in the first two cases in which the signal at the L terminal varies from 85% to 15% with a change in coupling length of just 10 µm. In the largest coupler, a more balanced distribution is obtained between both terminals, 55%. The directionality of the signal depends essentially on the Josephson-like oscillations and can be selected at will, demonstrating the functionality of these couplers as light-matter splitters, enabling a wide range of signal ratios just by changing one parameter.

To support the experimental findings, we performed numerical simulations of the polariton's propagation in coupled waveguides of the same shape as those used in the experiments. As explained in the introduction, to describe the propagation of polaritons we will use a coherent pump that creates polaritons with the same properties as those in the experiment. We start with the GPE *scalar* case, which does not consider the polarization of the polariton waves. In this case, the dynamics can be described by an equation for the slow varying amplitude of the polariton mode (order parameter function) $\psi$ [46, 47]

$$i\hbar\partial_t\psi = \frac{\hbar^2}{2m_{eff}}\nabla^2\psi - i\frac{\hbar}{2}\gamma\psi + V\psi + g_c|\psi|^2\psi + f(r,t) \quad (1)$$

where $m_{eff}$ is the effective mass of polaritons, $V$ is the potential trapping the polaritons, $\gamma$ describes coordinate-dependent losses, $g_c$ addresses the strength of the nonlinear (depending on the density) shift of the polariton frequency (chemical potential). The system is pumped by a monochromatic spatially localized coherent drive $f$.

The effect of the waveguide is modeled by an attracting potential $V$, trapping the polaritons in a potential well with the shape of the experimental waveguide. The polaritons are well but not completely localized inside the waveguides, therefore in the region where the waveguides go parallel and close to each other polaritons can interact through their evanescent fields (coupling region). In this way, the polaritons can tunnel from one waveguide to another. The depth of the potential is adjusted to reproduce the inter-waveguide tunneling period observed in the experiments. The effective mass of polaritons $m_{eff}$ was fitted using the experimentally measured dispersion of the polaritons [24].

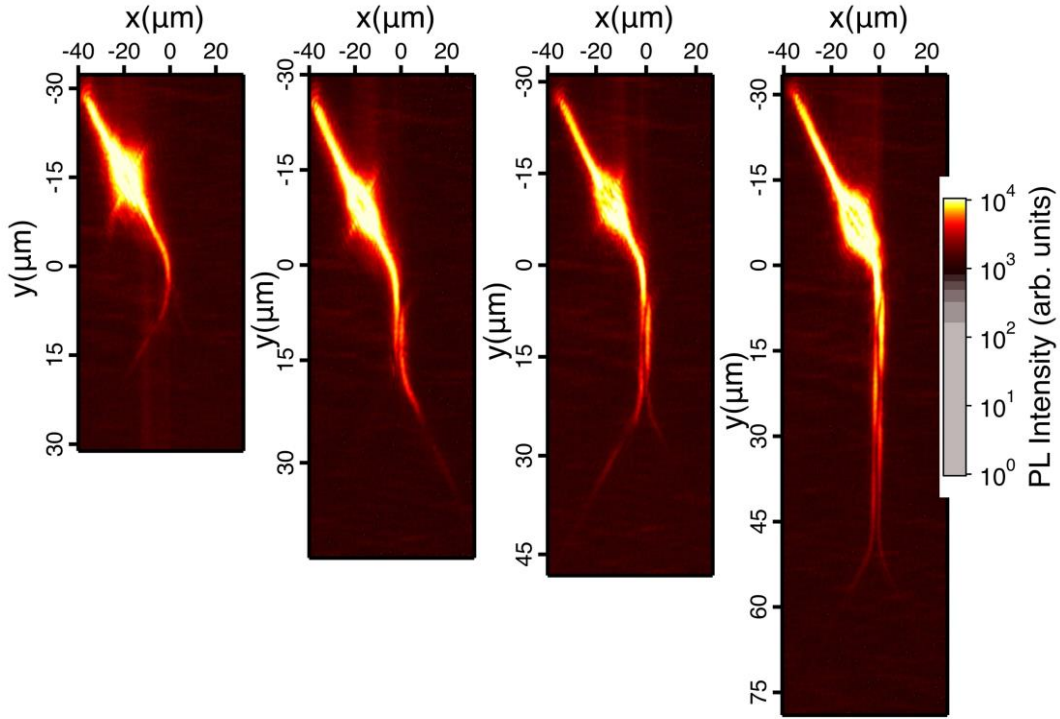

Figure 3.- Real-space emission maps of several couplers with different coupling lengths; from left to right <2, 10, 20 and 50 μm. L and R label the left and right output terminals respectively. The false colour logarithmic scale is saturated in the pump area to enhance the visibility of weaker signals at the output terminals.

For the numerical simulations, we extract $m_{eff}$ and $g_c$ from a fit of the dispersion relations and the blue shift of the polariton frequencies, respectively, obtaining $m_{eff} = \frac{\hbar^2}{0.55\ meV\cdot\mu m^2}$ and $g_c = 1\cdot 10^{-3}$ meV·μm². We also determine the decay length $L_d$ of the polaritons propagating in the waveguides, which allows us to find an approximate value of $\gamma = v_g/L_d = 0.24\ ps^{-1}$, where $v_g$ is the group velocity. It is reasonable to assume that the polariton losses are much smaller inside the waveguides than outside, so we take $\gamma$ to be 10 times lower inside. To emulate the experiment, we took the monochromatic pump $f = f_0(\vec{r})\exp(i\omega_p t)$ with the frequency $\omega_p$ fitted to the experimentally observed propagating polaritons. The specific spatial distribution of the pump described by function $f_0(\vec{r})$ is not relevant for our discussion; we took it in the form of narrow bell-shaped function situated at the axis of the input terminal $f_0(\vec{r}) = a_{ch}\exp(-\frac{|\vec{r}-\vec{r}_{ch}|^2}{w_{ch}^2})$ where $w_{ch}$ is the spatial aperture of the coherent drive (in the simulations presented in this paper the coherent pump is chosen to be $w_{ch} = 0.26\ \mu m$), $\vec{r}_{ch}$ is the radius vector defining the position of the pump centre and $a_{ch}$ is the amplitude of the pump, chosen to be small enough to ensure that nonlinear processes do not affect the polariton dynamics.

Let us start by discussing the dependence on the length of the coupling area of the polariton fractions guided to the left and right output terminals of the couplers. The results are summarized in Fig. 4 being a counterpart to experimental Fig. 3, which displays the total emission intensity (not polarization-resolved) along the couplers of different lengths. The polaritons created at the pump spot in the left input terminal propagate towards the bending area (not shown). At the bend the polaritons are redistributed entering the coupling region propagating along both waveguides (Fig. 4 (a1-a4).

Along the polariton propagation, there is a decay of their emission, due to different contributions such as polariton decay, microcavity losses and transfer efficiency between the arms. We have quantified this decay by plotting the integrated – over the width (w) – emission on the left ($E_L$) and right ($E_R$) coupler arm as a function of the propagation distance panel in Fig. 4 (b), showing a reduction of the signals of about 3 orders of magnitude.

The quasi-periodic bouncing of polaritons between the coupler arms seen in Fig. 4 (a1-a4), with a period of about 20 μm, which is close to that of the experiments, is clearly brought out in panel (b) in the coupling region, showing alternating maxima and minima for $E_L$ and $E_R$. This bouncing can be explained by the coupling between the waveguides [25]. As a result of these oscillations, the fraction of polaritons going into the right or the left output terminals, $R = \frac{I_L}{I_L + I_R}$, depends strongly on the length of the coupling area, $L_C$, as shown by our numerical results (*) in Fig. 4(c). $I_{L/R}$ are obtained integrating $E_{L/R}$ along the output terminals for different $L_C$. $R$ oscillates with a period very close to the period of the inter-waveguide bouncing. It is important to notice that this ratio does not reach zero or unity.

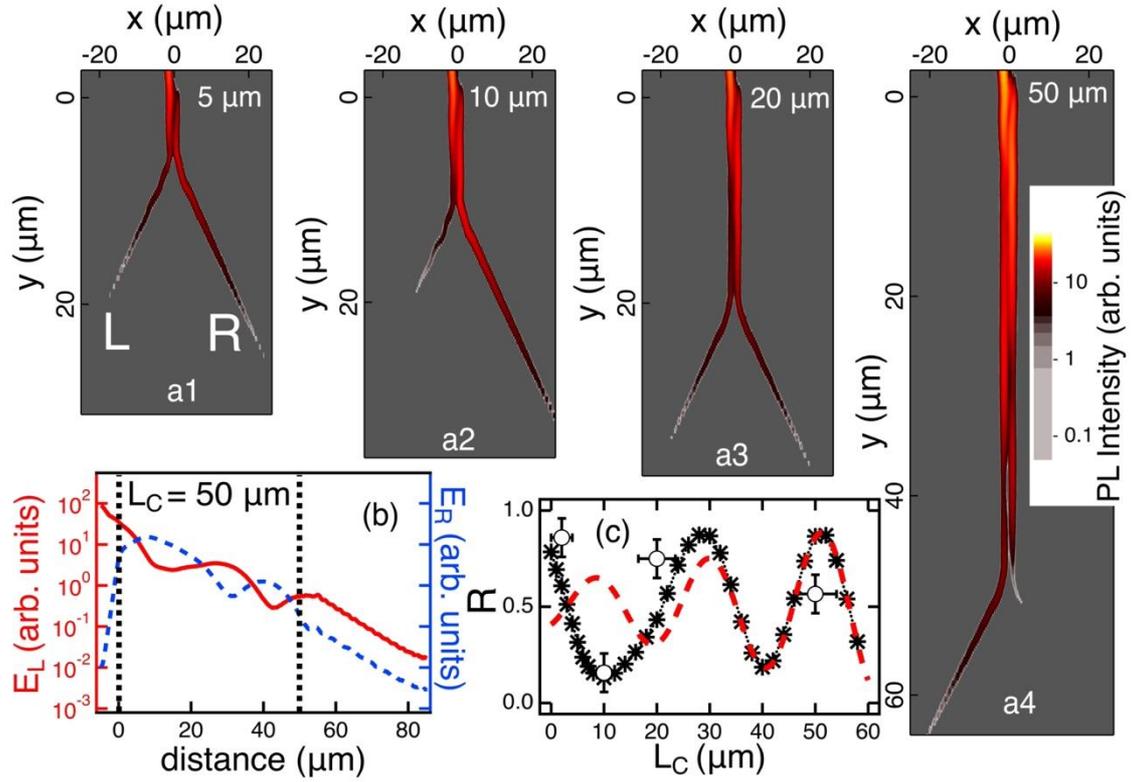

Figure 4.- (a1-a4) Numerically simulated real-space emission (not polarization resolved) maps of couplers with different coupling lengths; from left to right 5, 10, 20 and 50 μm. L and R label the left and right output terminals respectively for different coupling lengths. (b) Numerically simulated emission integrated over the width (w) on the left ($E_L$, solid red) and the right ($E_R$, dashed blue) coupler arm as a function of the propagation distance. The dotted vertical lines delimit the 50 μm long coupling region. (c) Numerically simulated ratio $R = \frac{I_L}{I_L+I_R}$ as a function of the coupling length, $L_C$ (*); $I_{L/R}$ denotes the PL intensity integrated in the left/right output terminal. Open dots compile the experimental ratio for four different couplers. The dashed red line corresponds to a fit to Eq. (2).

To understand this last fact, one must consider the beating of the modes accounting for their finite propagation length using a simplified analytical description, disregarding the polarization effects and the polariton higher order modes in their propagation along the couplers. The details are given in the supporting information, the $R_{an}(y = L_C)$ obtained therein [see Supporting information II, Eq. (S4)]

$$R_{an}(y) = \frac{1}{2} + \frac{|\alpha|\cos(\Delta k \cdot (y-y_{max}))}{e^{\Delta \gamma \, y}+|\alpha|^2 e^{-\Delta \gamma \, y}} \qquad (2)$$

renders the cosine dependence shown as a red dashed in Fig. 4 (b) with an inter-waveguide polaritons bouncing length of $\sim 20$ μm. $\Delta k$ / $\Delta \gamma$ are the differences between the wavevectors / losses of the propagating symmetric and antisymmetric polariton eigenmodes, and the two parameters $\alpha$ and $y_{max}$ depend on the distribution of the polaritons between the waveguides just after the bending of the input terminal into the beginning of the coupling area (see Supp. Inf. II for more details). These analytical results are appropriate for $L_C \gtrsim 30 \, \mu m$, where the contribution of higher order modes of the polariton can be disregarded (see Supp. Inf. III for more details).

## (B) Linear polarization

It has recently been shown that the polarization of the eigenstates in quasi one-dimensional polaritonic structures is linear [51, 52], and that they are pinned to the longitudinal and transverse axis of the device. Due to the geometry of these couplers, there are four eigenmodes: parallel/perpendicular to the coupling length (we label them as V/H, for vertical/horizontal polarization, see Fig. 1) and parallel/perpendicular to the terminals (we label them as D/A for diagonal - 45°- and antidiagonal -135°- polarization, see Fig. 1). Because of the inter waveguide interaction, these eigenstates have different propagation constants, what leads not only to oscillations of the linear polarization degree but to more complex dynamics.

To address the dynamics of this interaction, we have selected a coupler with nominal $L_c$ = 50 μm and a 0.2 μm separation between the 2 μm wide arms. In this way we guarantee a sufficiently long coupling area for the interaction to take place, while maintaining a large enough polariton condensate population.

To analyze these new results, we generalize Eq. (1) to the equations for the order parameter functions $\psi_{r,l}$ of the right- and left circularly polarized polaritons:

$$i\hbar \partial_t \psi_{r,l} = \frac{\hbar^2}{2m_{eff}} \nabla^2 \psi_{r,l} - i\frac{\hbar}{2}\gamma\psi_{r,l} + (V + g_c|\psi_{r,l}|^2 + g_{cx}|\psi_{l,r}|^2)\psi_{r,l} +$$
$$\frac{\hbar^2 \Delta m_{eff}}{2m_{eff}^2}(i\partial_x \pm \partial_y)^2 \psi_{l,r} + f_{r,l}(r,t) \qquad (3)$$

This model additionally includes the polariton frequency shift due to polaritons with different polarizations and, more importantly, the TE-TM splitting. The former ($g_{cx}$ term) accounts for the nonlinear interaction between polaritons of orthogonal polarizations, for which we take a typical value of $g_{cx} = -0.1 g_c$. The latter, characterized by the difference of the effective mass $\Delta m_{eff} = m_\parallel - m_\perp$, appears due to the anisotropy of the polariton dispersions, with masses $m_\parallel$ and $m_\perp$ for the TE and TM modes, respectively (the eigenmodes are linearly polarized perpendicular or parallel to the propagation direction). $m_{eff} = \frac{m_\parallel + m_\perp}{2}$ and $\Delta m_{eff}$ are obtained from a fit of the experimental dispersions. The TE-TM splitting is the most important for us because it is responsible for the polarization oscillations of the polaritons.

As our results will show below, the integrated polariton populations at the output terminals for different linear polarization azimuths, φ, depend strongly on Lc. Therefore, given the uncertainty in the coupling length determination ($L_c$ = 50 ±3 μm), we performed numerical simulations for different coupling lengths. These populations are defined, in the basis of linear polarizations, as $I_{L,R}(\varphi) = \iint |\psi_x|^2 \, dS \cos^2\varphi + \iint |\psi_y|^2 \, dS \sin^2\varphi$, where φ is the angle between the x-axis and the direction of the linear polarization analyzer, integrating over either the left (L) or the right (R) output terminal. A detailed discussion of the simulations is given in the section D of the Supporting information. These populations correspond to the experimental linear-polarization resolved PL. Figure 5 summarizes the main results regarding the linear polarization dynamics in this coupler: panel (a) depicts the experimental integrated PL for a nominally 50 μm long coupler at both output terminals (L dashed line, R solid line) as a function of the linear polarization azimuth; the remaining panels (b to d) compile the numerical simulations for different $L_C$'s. Sinusoid-like oscillations are clearly observed: their amplitudes, as well as the relative intensities at the right and left terminals, together with their phases are strongly dependent on $L_C$.

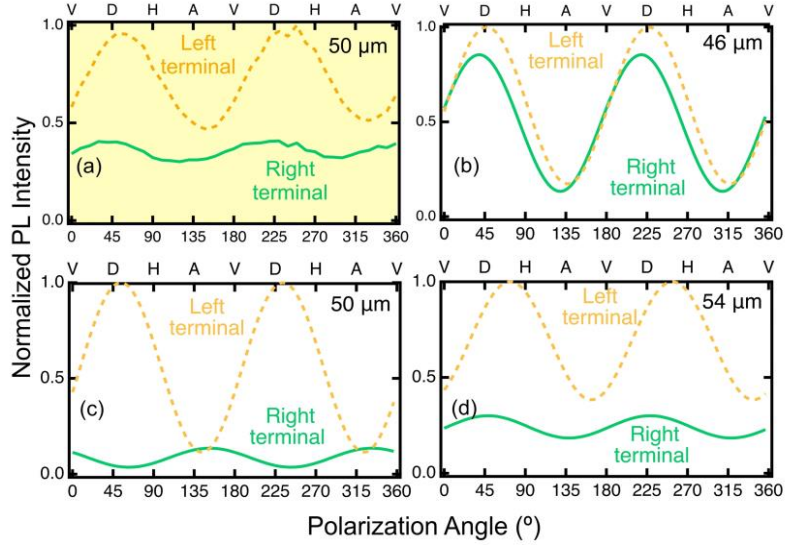

Figure 5.- Spatially integrated PL intensities at the output terminals (left dashed- and right solid-lines) as function of the polarization angle, measured with respect to the vertical, y axis. (a) Experiments for a coupler with $L_C$ =50 µm, d= 0.2 µm and w= 2 µm; (b) to (d) simulations for d= 0.2 µm, w= 2 µm and different coupling lengths, ranging from 46 to 54 µm.

Our results demonstrate that for Lc $\approx$ 50 µm, a larger population is routed towards the left terminal (Fig. 5c), while increasing (Fig. 5d) or decreasing (Fig. 5b) slighltly the coupling length significantly modifies the polariton distribution in both terminals and the relative phase between them, going from out of phase (Fig. 5c) towards in phase (Figs. 5b, d). The best agreement between our experimental results and the simulations is obtained for an $L_C$ of 54 µm slightly larger than our experimental one but within the uncertainty in the determination of the coupling length.

**IV. CONCLUSIONS**

We have presented a study of the transfer dynamics of polariton condensates between the arms of codirectional couplers, demonstrating the feasibility of directing polaritons towards different output terminals depending on the periodicity of Josephson-like oscillations, which can be tuned by selecting the appropriate coupling parameters, namely the coupling length and the spacing between the arms, demonstrating the functionality of these couplers as light-matter splitters. Here, we build upon our previous work by leveraging the periodic nature of these oscillations, along with variations in the coupling length, to select the polariton distribution along a specific output terminal. Future work dealing with polariton flow re-routing could be achieved by means of additional light beams impinging on the arms along the condensate's trajectory, or applying external magnetic fields, unique features for polariton couplers compared with photonic ones. The experiments are well accounted by simulations that are based on the generalized vector Gross-Pitaevskii equation using a coherent, resonant pump that replicates the polariton propagation. The oscillatory behavior of the fraction of polaritons entering a given output terminal is explained by a simplified analytical description, neglecting polarization effects and polariton higher order modes in their propagation along the couplers. The developed analysis reveals that the oscillatory dependency is only quasi-periodic because of the dissipative effects influencing the interference of the eigenmodes of the system. Analyzing the linear polarization of the emitted intensity at the output terminals of unexplored long couplers also obtains an oscillatory behavior as a function of the analyzer angle, which is well described by our model when the differences in propagation wavevectors and decay lengths of the different modes are taken into account.

**SUPPORTING INFORMATION**

The supporting information provides further details about the use of resonant excitation in the simulations, the polariton population splitting ratio between the waveguide's arms, the role of the high order modes on the polariton propagation and the polarization dependence of the integrated emission at the output terminals.

**ACKNOWLEDGEMENTS**

This work has been partly supported by the Spanish MINECO Grant PID2020-113445GB-I00. The work of AY was supported by Priority 2030 Federal Academic Leadership Program and Goszadanie no. 2019-1246.The Würzburg group acknowledges financial support by the German Research Foundation (DFG) under Germany's Excellence Strategy–EXC2147 "ct.qmat" (Project No. 390858490) and is grateful for support by the state of Bavaria. SK is grateful for support by the DFG under project KL3124/2-1.